\documentstyle[12pt]{article}
\begin{document}
\centerline{\bf Rayleigh scattering in rare-gas liquids}\vspace{1cm}

\centerline{G.M. Seidel, R.E. Lanou and W. Yao}
\vspace{.5cm}

\centerline{Physics Department, Brown University, Providence, RI 02912, USA}\vspace{.5cm}

\centerline{\bf Abstract}\vspace{.5cm}

The Rayleigh scattering length has been calculated for rare-gas liquids in the ultraviolet for the frequencies at which they luminesce. The calculations are based on the measured dielectric constants in the UV in the gas phase, in some cases extrapolated to the wavelength of luminescence. The scattering length may place constraints on the design of some large-scale detectors, using UV luminescence, being proposed to observe solar neutrinos and dark matter. Rayleigh scattering in mixtures of rare-gas liquids is also discussed.
\vspace{.5cm}

 \vspace{.5cm}

\vspace{4cm}

\newpage

\centerline{\bf I. Introduction}\vspace{.5cm}

Recent advances in particle/astrophysics have created the need for low-background, large-mass particle detectors capable of measuring very small energy depositions. A number of novel detectors have been proposed to meet the requirements of the search for dark matter in the laboratory, the measurement of the spectrum of p-p neutrinos from the Sun and the observation of neutrinoless double beta decay. Many of the proposals involve the use of rare-gas liquids, as this class of materials offers several advantages as large-mass, low-background absorbers. Helium \cite{us}, neon\cite{McKinsey} and argon\cite{icarus} are being investigated as detectors of solar neutrinos, and xenon is being studied in a number of laboratories as a detector for neutrinos\cite{xenon}, dark matter\cite{dark}, $\gamma$-rays\cite{gamma} and double beta decay\cite{2beta}. Liquid helium is also being used to measure the lifetime of the neutron\cite{lifetime} and is proposed as a medium in which to measure the electric dipole moment of the neutron\cite{dipole}. Both liquid  krypton\cite{krypton} and liquid xenon\cite{ypsil} have been developed for high-energy calorimetry in accelerator-based experiments. Being inert cryogenic fluids, the liquefied rare gases can be purified to levels not easily achieved in other materials. Furthermore, all the liquids are very efficient scintillators upon being ionized. They all fluoresce in the ultraviolet, radiatively decaying from an excimer state, a dimer formed by the binding of a ground state atom with an excited atom.  The final state is a pair of dissociated atoms. Helium luminesces in a broad band centered about 16~eV, while at the other end of the series, xenon luminesces at 7~eV. Xenon\cite{seguinot} and argon\cite{grosjean} have been measured to convert as much as 55\% of the kinetic energy of a primary ionizing particle into UV photons, Helium is somewhat less efficient, but still converts 35\% of the energy of a few hundred keV electron into scintillation\cite{us2}.

In all the liquids, the luminescence from the dimers that radiatively decay to the dissociated ground state has an energy of 1~eV or more below that of the first excited state of the atoms. As a consequence, the pure liquids, absent absorbing impurities, are transparent to their scintillation radiation. 
However, attenuation of luminescent radiation has been observed in the heavier liquids. This attenuation has been attributed to Rayleigh scattering from density fluctuations\cite{ishida}, but absorption by impurities has not been ruled out. Since several of the proposed detectors are dependent upon event location in the liquid to assist in background discrimination, Rayleigh scattering presents a potential design constraint. We explore below what can be inferred about the Rayleigh scattering length of the liquefied rare gases based on their measured physical properties.
\vspace{.5cm}

\centerline{\bf II. Scattering}\vspace{.5cm}

The well known expression for the inverse of the Rayleigh scattering length for a single component system can be found in Landau and Lifshitz\cite{landau}. It can be written in the form
\begin{equation} 
h = \frac{\omega^4}{6\pi c^4}\ \Big[ kT \rho^2  \kappa_T   \Big(\frac{\partial \epsilon}{\partial \rho}\Big)^2_T \ +\ \frac{kT^2}{\rho c_v}\Big(\frac{\partial \epsilon}{\partial T}\Big)^2_\rho\Big],
\label{ray}
\end{equation}
where the symbols have their usual meaning; $\omega$ is the angular frequency of the radiation, $c$ is the velocity of light, $k$ is Boltzmann's constant, $T$ is the temperature, $\rho$ is the liquid density, $\kappa_T$ is the isothermal compressibility, $c_v$ is the heat capacity at constant volume, and $\epsilon$ is the dielectric constant. 

The thermal properties of the rare-gas liquids appearing in Eq.~(\ref{ray}) are well measured. We have taken values of these quantities for Ne, Ar, Kr, and Xe from Ref.~\cite{rabin} and for He from Donnelly and Barenghi\cite{donn}. The values of density and compressibility used for the various liquids are listed in Table~I. The major difficulty in determining the scattering length arises in estimating the density dependence of the dielectric constant. Except for xenon there appear to be no measurements of the dielectric constants of the rare-gas liquids at the wavelengths at which they luminesce in the UV. Measurements that have been reported are (1) the refractive index in the UV of the rare gases at standard temperature and pressure, (2) the density dependence of the refractive index of the gases in the visible, and (3) the dielectric constant of the liquids in the visible. From these data we need to estimate $(\partial\epsilon/\partial\rho)_T$ for the liquid in the UV. 

In an atomic gas at sufficiently low densities such that interactions among atoms are of no consequence, the dielectric constant and the atomic polarizability of an isolated atom, $\alpha_0$, are related by the Clausius-Mossotti equation
\begin{equation}\frac{\epsilon -1}{\epsilon + 2} = \frac{4\pi}{3}\frac{N_a \alpha_0}{M} \rho,
\label{clausius}\end{equation}
where $N_a$ is Avogadro's number and $M$ is the atomic mass. The density, $\rho$, is expressed as mass per unit volume. At higher densities where interatomic interactions can influence the polarizability of an atom, Eq. (\ref{clausius}) is not correct. The dependence on density of the Clausius-Mossotti function, $(\epsilon -1)/(\epsilon + 2)$, is often expressed as a virial expansion,
\begin{equation} \frac{\epsilon -1}{\epsilon + 2} = A_\epsilon(\omega)\rho + B_\epsilon(\omega,T)\rho^2 + C_\epsilon(\omega,T)\rho^3 + \cdot \cdot\ ,
\label{virial} \end{equation}
in which the frequency and temperature dependence of the coefficients is explicitly noted. 

The dielectric virial coefficients have been measured for the rare gases by many groups. In particular, Achtermann, {\it et al.}\cite{achter} have made measurements at room temperature in the visible on all the gases up to densities close to one-half that of the liquids at their boiling points. The measured second and third dielectric virial coefficients are sufficiently small that the density dependence of the Clausius-Mossotti function can, for the purposes of these calculations, be treated as linear for all the rare-gases liquids. Furthermore, the temperature dependence of the second and third dielectric virial coefficients has been measured for the rare gases\cite{huot}. For He and Ar, the temperature variation of $\epsilon$ has been calculated\cite{mosz,koch} and compared to the existing data. The variation of the polarizability with temperature arises from the weak temperature dependence of the spatially averaged interactions among atoms. This effect is very small. Temperature fluctuations in the dielectric constant make a negligible contribution to the scattering length compared to the effect of density fluctuations. In Eq.~(\ref{ray}), the estimated values for the second term, involving $(\partial \epsilon/\partial T)_\rho$, are several orders of magnitude smaller than the first term, and are not considered further here. 

Since the contributions of the higher order virial terms are small at the liquid density, we write the Clausius-Mossotti function simply as \begin{equation}\frac{\epsilon(\omega) -1}{\epsilon(\omega) + 2} = A\ \rho
\label{last}
\end{equation}
To illustrate the accuracy of this approximation we list in Table~II the values of the parameter $A$ of Eq.~(\ref{last}) in the visible region of the spectrum for the rare gases measured at low density, calculated at the density of the liquid from the  measured coefficients of the virial expansion, Eq.~(\ref{virial}), and measured for the liquid. Note that $A$ measured for the liquids is, in all cases, no more than 1\% different from the measured value for the dilute gas. 

The refractive index of the rare gases Ne, Ar, Kr and Xe at STP has been measured by Bideau-Mehu, {\it et al.}\cite{bideau} in the UV between, roughly, 270 and 140~nm. They fit their data to either two-term or three-term empirical Sellmeier equations\footnote {There are misprints in Ref.~\cite{bideau} in the Sellmeier equations for neon and xenon, see Refs.~\cite{bulanin1,bulanin2}.}, which relate the index to oscillator strengths of the first one or two excited states and combine the influence of all the remaining states and the continuum into another term. Theoretical calculations of the dynamic polarization for the rare gases have been performed by a number of groups, the more recent computations being those of Bulanin and Kislyakov\cite{bulanin1,bulanin2}. The agreement between theory and experiment is good throughout the frequency range of the measurements. Differences are a few percent at worst.

The fact that in the visible the values of $A$ for the liquids and for the gases at STP are so close indicates that intermolecular interactions have very little effect on the atomic polarizability at the density of the liquid. 
Based on this agreement in the visible, we assume that the dielectric constants at the scintillation wavelength of the rare-gas liquids can be obtained from those determined in the gas. While the dielectric constants of xenon and krypton gas have been measured at the wavelengths for which they scintillate, those of helium, neon and argon have not. For neon and argon we used extrapolations of the data of Bideau-Medu, {\it et al.}\cite{bideau} together with the calculations of Bulanin and Kislyakov\cite{bulanin2} for the dynamic polarizability to obtain values of the dielectric constants at the wavelength of scintillation. In the case of helium, experimental data\cite{smith,abjean} extend only to 170~nm and are characterized by relatively large uncertainties.
The measurements of Smith, {\it et al.}\cite{smith} have been extrapolated in accordance with the calculations of Ref.~\cite{bulanin2}.

The Rayleigh scattering length of the various liquids are calculated using the first term of Eq.~(\ref{ray}). The thermodynamic parameters are listed in Table~I. From the values of $\epsilon$ listed in Table~III we compute the density variation of the dielectric constant from
\begin{equation} \Big(\frac{\partial \epsilon}{\partial \rho}\Big)_T = \frac{3A}{(1-A \rho)^2}= \frac{(\epsilon -1)(\epsilon +2)}{3\rho}.\end{equation}
The contribution of the density dependence of $A$ to $(\partial \epsilon/\partial \rho)$, as discussed above, is small and neglected. The results of the calculation for the scattering lengths are given in Table~III along with the results of the measurements for the three dense liquids by Ishida\cite{ishida} and others\cite{braem,chepel,akimov}. The calculations have been performed for all the liquids at their normal boiling points and, in addition, for helium at 0.1~K, the proposed operating temperature of a detector of solar neutrinos\cite{us}. 

The major uncertainty in calculating the scattering length arises from the determination of the dielectric constant. An error in $\epsilon$ of 1\% translates into an error in the scattering length of 25\% in the case of liquid helium but only to 4\% for xenon at the other end of the series. The uncertainty in the dielectric constant of the liquids results from possible errors due to extrapolation to the scintillation frequency and due to an unknown dependence on density. As a result, the calculated scattering length is judged to have an uncertainty of a factor of 2 in the case of helium , 40\% for neon, 35\% for argon, 25\% for krypton and 20\% for xenon. 
The calculated  attenuation lengths agree, in the three cases where a comparison can be made, with the experimentally measured values, especially given the uncertainty in the measured results as represented by the spread in values. The agreement  suggests that the measured attenuation in the rare-gas liquids is due to Raleigh scattering and is not the consequence of absorption by impurities.

Xenon is the one liquid the dielectric properties of which have been measured in the UV. Barkov, {\it et al.}\cite{barkov} measured the refractive index, $n$, of liquid Xe at the luminescent frequency to be $n=1.565\pm.01$, while Lopes\cite{lopes} reports the value of $1.72\pm.02$. The stated precision of each of the measurements is far less than the difference between the two values. We note that these measured values for the refractive index bracket the value obtained from the calculated dielectric constant, $n = \sqrt{\epsilon} = 1.69$. The value of $\epsilon$ for the liquid we have deduced from the optical data on the gas is in reasonable agreement with the refractive index quoted by Lopes.  \vspace{.5cm}

\centerline{\bf III. Mixtures}\vspace{.5cm}

One method by which it is possible to circumvent the short scattering length of luminescent light of a material is to introduce small quantities of a second soluble component that fluoresces at a longer wavelength. The solvent remains the absorber of the detector. When a particle is stopped, the energy within the ionized and excited atomic system of the solvent is transferred to excimers involving the solute, leading to the emission of radiation at a longer wavelength. Since the Rayleigh scattering length is proportional to the fourth power of $\lambda$, the effect on the scattering can be dramatic. 

There is one complication, however. In a multicomponent system, fluctuations in the concentrations of the constituents lead to spatial variations in the dielectric constant and to an increase in scattering\cite{kirkwood}. For a two-component system an additional term must be added to Eq.~(\ref{ray}) of the form
$$\frac{\omega^4}{6\pi c^4}\ \Big[\frac{kT}{n_1(\partial \mu_2/\partial x)}\big(\frac{\partial \epsilon}{\partial x}\big)^2\Big], $$
where $x$ and $\mu_2$ are the concentration and chemical potential of the solute atoms, respectively, and $n_1$ is the number density of the solvent atoms. If the solute is dilute and can be treated as an ideal gas, then $\mu_2 = \mu^0_2 +kT\ ln(x)$~\cite{landau2}, and the term becomes 
$$\frac{\omega^4}{6\pi c^4}\ \Big[\frac{x}{n_1}\big(\frac{\partial 
\epsilon}{\partial x}\big)^2\Big]. $$
Using this approximation, and $\epsilon\approx x\epsilon_1 +(1-x)\epsilon_2$, we have calculated the Rayleigh scattering length for the two mixtures measured by Ishida\cite{ishida}. The results are listed in Table~IV. In both cases with a 3\% concentration of Xe, the term in $(\partial\epsilon/\partial x)$, rather than the term in $(\partial \epsilon/\partial \rho)$, is the principal contributor in determining the scattering length. Again, the agreement between the calculated values and the experimental results appears reasonable, especially given the variation in the length observed in separate measurements.\vspace{.5cm}

\centerline{\bf IV. Discussion} \vspace{.5cm}

Except for helium below 1~K, Rayleigh scattering from density fluctuations in the rare-gas liquids poses a potential constraint on some designs of large detectors using UV scintillation from these cryogenic fluids. The calculated scattering lengths are in reasonable agreement with the measurements that are available. 

One means of circumventing the short scattering length in the pure liquids is to add to the liquid a solute that shifts the scintillation to longer wavelength. The addition of Xe to Ar is a good example. In the case of liquid xenon, a material of considerable interest for use in large detectors, the scattering length is short, and there is no rare gas which as a solute can shift the radiation to longer wavelengths. However, there are other possible options. Many molecules besides the rare gases are soluble in liquid xenon. For example, the addition of fluorine to liquid xenon\cite{jara} results in emission in three bands in the UV with wavelengths centered at 193, 248 and 351~nm. These emissions are the result of the radiative decay of excimers, the XeF$^*$ dimer and Xe$_2$F$^*$ trimer, to their respective ground states. Fluorine has the advantage, in low background experiments, that it has no natural or cosmogenic long-lived radioactive isotopes. The efficiency of the energy transfer from Xe to mixed excimers and the dependence of this process on concentration of the solute must be explored to assess whether the addition of fluorine is a viable option.  Furthermore, pulse shapes need to be studied to determine if it is possible, with the use of a solute to shift frequency, to discriminate an energy deposition by a nuclear recoil from that by an electron, a feature that is essential in some types of detectors. In Xe:Ar mixtures\cite{kubota, conti} the time dependence of the pulse is substantially changed from that in either pure Ar or Xe, as is to be expected. The pulse shape is dependent, in part, on the processes involved in the transfer of energy from the solvent to the solute. In a pure liquid such transfer processes are not required.
\vspace{.5cm}

\centerline{\bf V. Acknowledgments} \vspace{.5cm}

We are grateful to Dr. I. Lopes (University of Coimbra) for providing results of the dielectric constant of liquid xenon and to Professor H.J. Maris for helpful comments. We also thank the referee for a careful reading of the manuscript and pointing out a misprint in Ref.~\cite{bideau} that we had overlooked. This work was supported by the DoE, grant DE-FG02-88ER40452.

\newpage
\vspace{.5cm}
    
\centerline{
\begin{tabular}{lccc} 
\multicolumn{4}{c}{TABLE I. \ Properties of liquefied rare gases.}\\  \hline \hline
liquid &boiling & $\rho$&$\kappa_T\times 10^{10}$\\
& point &density &compressibility \\  
& K &g~cm$^{-3}$  &  cm$^2$~dyne$^{-1}$  \\ \hline
He &4.2&0.125&208  \\
Neon&27.1 &1.205 &4.95\\
Argon&87.3 &1.39&2.18 \\
Krypton&120&2.41&1.86\\
Xenon& 165&2.94&1.68\\
 \hline \hline
\end{tabular}  }\vspace{.5cm}

\newpage
\vspace{.5cm}

\centerline{
\begin{tabular}{lccc} 
\multicolumn{4}{c}{TABLE II. $A$ (cm$^3$/g) in the visible for liquefied rare gases.}\\
\hline \hline
liquid &gas$^{\rm a}$& gas extrapolated$^{\rm a}$ &liquid \\   
& low density&to liquid density &at boiling pt.\\ \hline
He &0.1303 &0.1297 &0.1291\\
Neon&0.0496 &0.0490 &0.0490$^{\rm b}$\\
Argon&0.1050 &0.1039&0.1045$^{\rm c}$\\
Krypton& 0.0765&0.0758&0.0771$^{\rm c}$\\
Xenon&0.0788 &0.0763&0.0801$^{\rm c}$\\  \hline \hline
\multicolumn{4}{l}{$^{\rm a}$\scriptsize Calculated using data from Ref.~\cite{achter}.}\\
\multicolumn{4}{l}{$^{\rm b}$\scriptsize From Ref.~\cite{pashkov}. Data taken in the rf, corrected to the visible.}\\
\multicolumn{4}{l}{$^{\rm c}$\scriptsize From Ref.~\cite{sinnock}.}\\
\end{tabular}  }\vspace{1cm}

\newpage
\centerline{
\begin{tabular}{lcccc} 
\multicolumn{5}{c}{TABLE III. Rayleigh scattering length for liquefied rare gases.}\\
\hline \hline
liquid &scintillation&dielectric&scattering length &scattering length \\  
 &wavelength&constant &calculated&measured\\   
&nm && cm &cm\\ \hline
He at 4.2 K&78    &1.077$^{\rm a}   $&$600 $            & \\
He at 0.1 K&78    &1.089$^{\rm a}   $&$2\times10^4$  & \\
Neon       &80    &1.52$^{\rm b}    $& $60  $        & \\
Argon      &128   &1.90$^{\rm b}    $&$90 $          & 66$^{\rm d}$ \\
Krypton    &147   &2.27$^{\rm c}    $&$60 $           & 82$^{\rm d}$, 100$^{\rm g}$\\
Xenon      & 174  &2.85$^{\rm c}    $&$30$           & 29$^{\rm d}$, 40$^{\rm e}$, 50$^{\rm f}$\\  \hline \hline
\multicolumn{5}{l}{$^{\rm a}$\scriptsize From Ref.~\cite{smith} extrapolated to 78~nm according to Ref.~\cite{bulanin2}.}\\
\multicolumn{5}{l}{$^{\rm b}$\scriptsize From Ref.~\cite{bideau} extrapolated to the scintillation wavelength according to Ref.~\cite{bulanin2}.}\\
\multicolumn{5}{l}{$^{\rm c}$\scriptsize From Ref.~\cite{bideau}.}\\
\multicolumn{5}{l}{$^{\rm d}$\scriptsize From Ref.~\cite{ishida}.}\\
\multicolumn{5}{l}{$^{\rm e}$\scriptsize From Ref.~\cite{braem}.}\\
\multicolumn{5}{l}{$^{\rm f}$\scriptsize From Ref.~\cite{chepel}.}\\
\multicolumn{5}{l}{$^{\rm g}$\scriptsize From Ref.~\cite{akimov}.}\\
\end{tabular}  }\vspace{1cm}
\newpage
\centerline{
\begin{tabular}{lcccc} 
\multicolumn{5}{c}{TABLE IV. Rayleigh scattering length for mixtures of liquefied rare gases.}\\
 \hline \hline
liquid &scintillation&dielectric&scattering length &scattering length \\  
 &wavelength&constant&calculated&measured\\   
&nm && cm &cm\\ \hline
3\% Xe in Ar&174&1.29&280& 170,118$^{\rm a,b}$\\
3\% Xe in Kr&174&1.97& 170&136$^{\rm a}$\\  \hline \hline
\multicolumn{5}{l}{$^{\rm a}$\scriptsize From Ref.~\cite{ishida}.}\\
\multicolumn{5}{l}{$^{\rm b}$\scriptsize Two separate measurements.}\\
\end{tabular}  }\vspace{1cm}

\end{document}